\documentclass[aps,prb,twocolumn,groupedaddress,noshowpacs]{revtex4}
\usepackage{graphicx}
\begin{document}
\title{Dielectric response of modified Hubbard models with neutral-ionic
  and Peierls transitions}

\author{Zoltan G.~Soos, Sharon A.~Bewick}
\affiliation{Department of Chemistry, Princeton University, Princeton}
\author{Andrea Peri, Anna Painelli}
\affiliation{Dip. di Chimica GIAF
  Universit\`{a} di Parma,
and INSTM-UdR Parma, 43100 Parma, Italy}

\date{\today}

  \begin{abstract}

The dipole $P(F)$ of systems with periodic boundary conditions
(PBC) in a static electric field $F$ is applied to one-dimensional Peierls-Hubbard models for organic
charge-transfer (CT) salts. Exact results for $P(F)$ are obtained for
finite systems of $N $= 14 and 16 sites that are almost converged to
infinite chains in deformable lattices subject to a Peierls transition.
The electronic polarizability per site,
$\alpha_{el} = (\partial P/\partial F)_0$,
of rigid stacks with alternating transfer integrals $t(1 \pm \delta)$
diverges at the neutral-ionic transition for $\delta$ = 0 but remains
finite for $\delta > 0$ in dimerized chains. The Peierls or dimerization
mode couples to charge fluctuations along the stack and results in large
vibrational contributions, $\alpha_{vib}$, that are related to
$\partial P/\partial \delta$ and that peak sharply at the Peierls transition.
The extension of $P(F)$ to correlated electronic states yields
 the dielectric response $\kappa$ of models with neutral-ionic
or Peierls transitions, where $\kappa$ peaks $>100$ are found with parameters
used previously for variable ionicity $\rho$
 and vibrational spectra of CT salts. The calculated $\kappa$ accounts
 for the dielectric response of CT salts based on substituted TTFs
(tetrathiafulvalene) and substituted CAs (chloranil).
The role of lattice stiffness appears clearly in models:
soft systems have a Peierls instability at small $\rho$
 and continuous crossover to large $\rho$, while stiff stacks such as
TTF-CA have a first-order transition with discontinuous $\rho$
 that is both a neutral-ionic and Peierls transition.
The transitions are associated
with tuning the electronic ground state of
insulators via temperature or pressure
in experiments, or via model parameters in calculations.
\end{abstract}

\maketitle
\section{Introduction}
Torrance et al. \cite{nit}
reported the first neutral-ionic transition (NIT) in
the organic charge-transfer (CT) salt TTF-CA. Tetrathiafulvalene (TTF)
is a potent $\pi$-electron donor (D), while chloranil (CA) is a strong
$\pi$-acceptor (A). As confirmed subsequently by detailed vibrational and
  structural studies,\cite{ttfca,ttfcaj,ttfcar}
  the 300 K structure contains mixed regular stacks
..D$^{\rho +}$A$^{\rho -}$D$^{\rho +}$A$^{\rho -}$...
  with face-to-face overlap and ionicity $\rho \sim$ 0.3. Equal
intermolecular interaction with each neighbor follows from inversion
symmetry at the centers of TTF and CA. Below the first-order transition at
T$_c$ = 81  K, the dimerized ionic stacks have  $\rho \sim$ 0.7
and alternately short and long intermolecular spacing. The ionic
  lattice is unstable to a transition that becomes a spin-Peierls
transition \cite{spin} in the limit $\rho \rightarrow$ 1.  In TTF-CA and other
  CT salts with discontinuous $\rho$ at T$_c$, the neutral-ionic and Peierls
transitions coincide.

Systems with neutral-ionic or valence transitions are necessarily rare:
two ground states (GS) with different charge densities must be almost
degenerate in order to switch between them by changing temperature or
pressure. Horiuchi et al. \cite{hprb,hprl,hjpsj,hjacs,okimoto,science}
have recently studied NITs in a series of
substituted TTFs and CAs whose static dielectric responses have large
peaks ($\kappa >$ 100). Their interpretation is a quantum phase
transition between a neutral-regular and an ionic-dimerized GS controlled
by an external parameter such as temperature or pressure. \cite{science}
No modeling of
$\kappa$ is proposed, however, and no modeling is possible without
the static polarizability of extended systems. In this paper we compute
the static polarizability of Peierls-Hubbard models for organic CT salts
and vary GS parameters to induce transitions. We obtain explicitly
electronic and vibrational contributions to $\kappa$
for correlated states. Systems with continuous $\rho$ at the $\kappa$
  peak and $\rho <$ 1/2 throughout are of particular interest,
since they do not fit the simple paradigm of neutral-regular or
ionic-dimerized.

The dielectric constant, $\kappa$, is the GS polarizability $\alpha$,
or induced dipole moment, per unit cell in the crystal.
The modern theory of polarization in insulators \cite{ksv}
has resolved the problem
of a position or dipole operator in systems with periodic boundary
conditions (PBC). Resta \cite{resta} introduced the following
expression for the expectation
value of the dipole moment, $P$:
\begin{equation}
   P=\frac{1}{2\pi} Im \ln \langle \psi |\exp{(i\frac{2\pi M}{N})}|\psi\rangle
\equiv \frac{1}{2\pi}  Im \ln Z
\label{p}
\end{equation}
Here $|\psi\rangle$ is the exact electronic GS, $M$ is the conventional
dipole operator and PBC are ensured via supercells. Polarization
in insulators is related to the {\it phase} rather than the {\it amplitude}
  of $\psi$
  and can be formulated as a Berry phase. $P$ applies to correlated as well as
noninteracting systems and provides a unified approach to previous treatments
of polarization and GS localization in insulators.\cite{restaloc}
  The magnitude of $Z$ is
finite in insulators and zero in conductors. The phase or twist
operator in Eq. (1) also appears in other contexts. \cite{twist}
We will work with $P$ for the exact GS of
modified Hubbard models with $N$ sites.

Although a static electric field $F$ is not compatible with PBC,
Nunes and Gonze \cite{nunes}
showed recently how to combine $F$ and $P$. They propose to
define $P(F)$ with $\psi(F)$ in Eq.~(\ref{p})
  and to minimize a functional that,
in a one-dimensional system, is
\begin{equation}
   E(\psi_F,F)=E(\psi_F)-NFP(\psi_F)
\label{energyfunctional}
\end{equation}
Several groups \cite{souza,umari,souza2,rignanese}
have applied Eq.~(\ref{energyfunctional})
  to the electronic dielectric constants of
metal oxides and silicates, using Wannier functions for $\psi(F)$
and perturbation theory
to find corrections to $\psi(0)$. As shown below, correlated models with
finite $N$ are simpler because the basis is finite, which allows exact
solution for  $\psi(F)$ in Eq.~(\ref{energyfunctional}).
The procedure in Section 2 differs from the
model-exact treatment of static or dynamic nonlinear optical
responses,\cite{soosnlo}
however, since the dipole operator $M$ is now in the exponent.

The optical phonon  of the mixed regular stacks
..D$^{\rho +}$A$^{\rho -}$D$^{\rho +}$A$^{\rho -}$...
describes a rigid displacement of either sublattice,
and in the hypothesis of linear electron-phonon (e-ph) coupling
leads to alternating CT integrals: $t(1+\delta)$, $t(1-\delta)$. We
assign the optical phonon a harmonic potential with frequency
$\omega_P$ at $\rho$ = 0.  With increasing $\rho$, e-ph coupling
generates large anharmonicities\cite{freo} and an overall
softening of the vibrational frequency, which vanishes at the Peierls
transition. A static electric field, $F$, strongly affects the
lattice degrees of freedom at intermediate $\rho$, especially near
the Peierls transition. The polarizability per site has then substantial
electronic and vibrational contributions,
\begin{equation}
   \frac{dP}{dF}= \frac{\partial P}{\partial F}+
\frac{\partial P}{\partial \delta}\frac{\partial \delta}{\partial F}=
\alpha_{el}+\alpha_{vib}
\label{alphaelalphavib}
\end{equation}
The $P$ derivatives above illustrate that physical processes are
associated with changes \cite{restaloc} of $P$
  and underscore the central role of $P$
  for the dielectric response. In the vibrational part,
$\frac{\partial P}{\partial \delta} $ is related to the IR
intensity of the Peierls mode\cite{freo}
  and $\frac{\partial \delta}{\partial F}$
to the field-induced change in dimerization. As anticipated from
non-interacting systems,\cite{freo}
  $\alpha_{vib}$ has a peak at the Peierls transition
that is further enhanced by the frequency softening.
 The electronic part, by contrast, diverges at the NIT
of the rigid regular lattice but varies slowly near the Peierls instability
of deformable lattices.

  The paper is organized as follows. Section 2 presents the exact solution of
$ P(F)$ using Eq.~(\ref{energyfunctional}) for quantum cell models with a
large but finite basis. The electronic polarizability of modified Hubbard
models with fixed alternation is obtained in Section 3. The roles of lattice
vibrations are introduced  in Section 4 and
developed explicitly for the dimerization mode and its softening at
the Peierls transition. Section 5 contains polarizability, ionicity and
related results for Peierls-Hubbard models with equilibrium dimerization.
The $\kappa$ peak at the Peierls transition is due to $\alpha_{vib}$
  and increases with the lattice stiffness. We discuss in Section 6 the
applicability of model results to dielectric peaks in organic CT salts.

\section{Static electronic polarizabilities}

Quantum cell models typically have a single Wannier orbital $\phi_p$
per site that, in principle, can incorporate intra-site correlations.
\cite{soosklein}
On-site repulsion $U > 0$ is kept in Hubbard models and long-range Coulomb
interactions in the Pariser-Parr-Pople model.\cite{sooshayden} Total spin $S$
is usually conserved, as in models for CT salts, but this condition
can be relaxed. The defining feature of cell models is a large but finite
many-electron basis in real space that is complete for a finite number of
orbitals $\phi_p$.  Slater determinants of $\phi_p$'s can be used in general
for eigenstates of $S_z$, but linear combinations of determinants are
required for the smaller basis when $S$ is conserved.
Easily visualized valence bond (VB) \cite{vb} diagrams are then convenient and
used below.

   The normalized singlet ($S = 0$) diagram $|k,q\rangle$ specifies
the occupation number $n_{pk}$ = 0,1 or 2 at each site $p = 1, 2, ... N$
  and the spin pairing of all singly-occupied site. The index $q$
  is introduced below for the eigenvalue $M_k$ of the dipole operator.
The VB basis is orthogonal with respect to charge, but there is overlap
$S_{kk'}$ between diagrams with identical $n_{pk}$ and
different spin pairing. The operator $M$ for $N$ sites with uniform
spacing $a = 1$ and $e = 1$ is
  \begin{equation}
  M=\sum_{p=1}^{N} p q_p=\sum_{p=1}^{N} p(z_p-n_p)
\label{emme}
  \end{equation}
Here $q_p = z_p - n_p$ is the charge operator, with $z_p$ = 0 at A
and $z_p$ = 2 at D, and there are two electrons per DA unit cell.
The neutral GS has regular spacing $a$ along the stack while the dimerized
ionic GS has alternating spacing $a\pm u$.  For simplicity we use $M$ in
Eq.~(\ref{emme}) for both regular and dimerized GS, thereby neglecting
the motion of $\pm e\rho$ charges; corrections for $\pm u$
  can readily be included.\cite{freo} $M$ is diagonal in the $|k,q\rangle$
basis. Its eigenvalues  are the integers, modulo $N$,
$M_k = 0, \pm 1, \pm 2, ..,-N/2$. We take them as $q = 2\pi M_k/N$ for
$-\pi \le q < \pi$ in the first Brillouin zone.
The matrix element in Eq.~(\ref{p}) is then\cite{freo}
   \begin{equation}
\langle \psi | \exp{i \frac{2 \pi M}{N}} | \psi \rangle =\sum_q
W_q \exp{iq} \equiv Z_x +Z_y
\label{expect}
  \end{equation}
where $W_q \ge  0$ is the weight of basis vectors $|k,q\rangle$ with
$q = 2\pi M_k/N$. Diagrams with $q \sim 0$ have greatest weight on the
neutral side and lead to $Z_x \sim 1$, while diagrams with $q \sim \pm \pi$
  and $Z_x \sim -1$ dominate on the ionic side.
The regular chain has real $Z$ by symmetry and $Z = 0$ at the NIT.

     Since the VB basis is complete, the GS can be expanded as
\begin{equation}
   \psi(F)=\sum_{k,q} c_{kq}(F)|k,q\rangle
\label{psioff}
\end{equation}
The linear coefficient $c_{kq}$ are real and known at $F = 0$
  by hypothesis. Substitution of Eq.~(\ref{psioff}) into
Eq.~(\ref{energyfunctional}) and varying the $c_{kq}$
  leads to a secular determinant. As discussed elsewhere,\cite{vb}
  sparse-matrix
methods that exploit completeness are much more convenient for large
( $> 10^6$) basis sets. We consequently seek an operator whose matrix
  elements correspond to $P(F)$ in Eq.~(\ref{energyfunctional}).
The partial derivative of $P(F)$ with respect to $c_{k'q}$ is
  \begin{equation}
    \frac {\partial P(F)}{\partial c_{k'q}} =\sum_{k,q} c_{k,q}(F) S_{k'k}
\left\{\frac{Z_x(F)\sin{q}-Z_y(F)\cos{q}}{2\pi \left[Z_x(F)^2
+Z_y(F)^2\right]}\right\}
\label{dpoff}
  \end{equation}
We define the quantity in $\{...\}$ as $\mu_{kq}$ and interpret it as the
induced dipole of diagram $|k,q\rangle$. Since precisely the same result
is obtained by introducing diagonal energies $-NF\mu_{kq}$
  for each diagram, the desired induced-dipole operator is
\begin{equation}
   \Delta M(F) =\frac{N}{2 \pi} Im \frac{exp{(i 2\pi \hat M /N)}}{Z(F)}
\label{deltap}
\end{equation}
It follows immediately that $|k,q\rangle$ is an eigenfunction of $\Delta M$
  with eigenvalue $N\mu_{kq}$, as given by Eq.~(\ref{dpoff}).
The large response or polarizability for small $Z(F)$ is also evident.

The minimization of the energy functional in Eq.~(\ref{energyfunctional})
  at finite $F$
  is equivalent to finding the GS of the Hamiltonian
  \begin{equation}
    H(F)=H-F\Delta M(F)
\label{hoff}
  \end{equation}
Given the action of $H(F)$ on any diagram, we obtain $\psi(F)$
  as usual \cite{vb} except for one complication. Since we know $Z(0)$
  from  Eq.~(\ref{p}), but need $Z(F)$ in Eq.~(\ref{deltap}),
  iterations are needed. We start with $\Delta M(0)$ to obtain $Z^{(1)}(F)$
  and hence $\Delta M^{(1)}(F)$, which is then used in Eq.~(\ref{hoff})
  to find $Z^{(2)}(F)$ and $\Delta M^{(2)}(F)$.
We repeat until $Z(F)$, $\psi (F)$ and $P(F)$ have converged.
The GS with $\Delta M(0)$ requires the same computational effort as
$\psi(0)$ and suffices for the linear response, $\alpha_{el} = P'(0)$.
More precisely, we use $\Delta M(0)$ in Eq.~(\ref{deltap}) to find
$\psi(F)$ at small $F$, compute $P(F)$ in Eq.~(\ref{p}) and evaluate
$ P'(0)$.  We have $\langle \psi(F)|\Delta M(0)|\psi(F)\rangle=
FP'(0)$ and, as seen directly from Eq.~(\ref{dpoff}),
  $\langle \psi(F)|\Delta M(F)|\psi(F)\rangle=0 $
The interpretation of Eq.~(\ref{deltap}) as
the induced-dipole operator is now clear:
$\Delta M(0)$ generates the linear correction to $P(F)$.

We note that
the Berry-phase formulation of $P(F)$
in Eq.~(\ref{energyfunctional})
differs from  standard
 expressions for finite systems with open boundary conditions.
In finite systems, the
interaction with a field is $H - F M$ instead of the PBC
Hamiltonian in Eq.~(\ref{hoff}).
The dipole operator $M$ is diagonal in the VB basis,
with  $M_k = \sum_p p q_{pk}$ for diagram $|k,q\rangle$.
Now $\Phi(F)$ is the GS of $H - FM$ and the electronic polarization
is $\langle \Phi(F) | M |\Phi(F)\rangle$.
  In contrast to PBC, there is no iteration and $M_k$ is not modulo $N$.
The theoretical ideas and computational difficulties that motivated the
PBC formulation of $P$ in Eq.~(\ref{p}) and
  its coupling to $F$ in Eq.~(\ref{energyfunctional}) do not arise
in quantum cell models, whose limited basis is unphysical in this respect.
But extrapolations of linear and nonlinear optical coefficients to infinite
systems have been widely discussed \cite{karna}
for both noninteracting (H\"uckel)
and correlated (Hubbard, Pariser-Parr-Pople) models, especially in
connection with frequency-dependent responses. The velocity operator
version of $M$ is often used to estimate transition dipole moments
 for extended (PBC) systems  and is consistent
with the real-space results. \cite{spano}
Susceptibilities obtained with the familiar sum-over-states (S0S)
approach require all excited states,
a difficult task for systems with large correlated
basis, but can be related to GS responses with correction vectors. \cite{soosnlo} The calculation of $P(F)$ and its
derivatives only requires the GS.

\section{ Electronic polarizability of modified Hubbard models}

         To illustrate the PBC polarizability of rigid lattices,
we consider Hubbard models that have been applied to the
  electronic structure of both organic CT salts
\cite{strebel,anu} and transition-metal oxides \cite{egami}.
  Face-to-face $\pi$-overlap in CT salts leads to transfer integrals $t$
  between sites with energy $-\Delta$ at D and $\Delta$ at A.
The modified Hubbard model is \cite{strebel}
\begin{eqnarray}
   H_{Hu}&=&-\sum_{p,\sigma} \left[ 1-\delta(-1)^p\right](a^\dagger_{p,\sigma}
a_{p+1,\sigma} + H.c.) \nonumber\\
&&+\sum_p \left[\Delta(-1)^pn_p+U\frac{n_p(n_p-1)}{2}\right]
\label{hhu}
\end{eqnarray}
We take $t = 1$ as the unit of energy, consider alternation $0 <\delta < 1$
along the chain, and assume equal on-site $U > 0$ for A and D sites.
Coulomb or other interactions can be added to $H_{Hu}$
  without increasing the correlated basis. Such models conserve $S$, have
$C_{N/2}$ symmetry and require comparable effort to obtain the exact GS.

         On physical grounds, high-energy D$^{2+}$ and A$^{2-}$
  sites can be neglected to obtain a restricted basis that is usefully smaller.
Formally, we define $\Gamma = \Delta- U/2$ and take the limit $\Delta,U
\rightarrow \infty$ in Eq.~(\ref{hhu}) while keeping finite $\Gamma$.
The electronic problem for rigid lattices simplifies to
\begin{eqnarray}
   H_0(\delta,\Gamma)&=& -\sum_{p,\sigma} \left[ 1-\delta(-1)^p\right]
(a^\dagger_{p,\sigma} a_{p+1,\sigma} + H.c.) \nonumber\\
&&+\sum_p \Gamma(-1)^pn_p
\label{h0}
\end{eqnarray}
The GS for $\Gamma >> 0$ is the neutral ($\rho \sim 0$)
lattice with $N$ electrons paired on $N/2$ donors, while the GS for
$\Gamma << 0$ is the ionic lattice ($\rho \sim 1$) with one electron per site
and overall singlet pairing. The regular ($\delta = 0$) stack has an
NIT as a function of $\Delta$ in Eq.~(\ref{hhu}) or $\Gamma$ in Eq.~(\ref{h0})
  that has been studied by different methods, \cite{anu}
  including exact GS energies and
$\rho$ up to $ N = 16$ for Eq.~(\ref{hhu}) and up
to $N=$ 22 for Eq.~(\ref{h0}).
\begin{figure}
\begin{center}
\includegraphics* [scale=1] {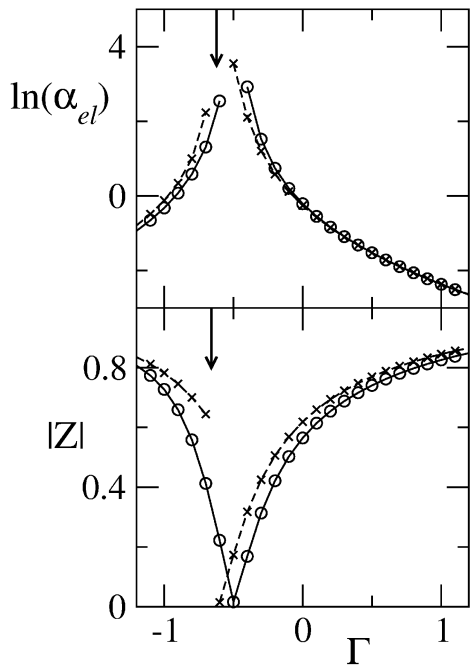}
\end{center}
\caption{The logarithm of the linear electronic susceptibility, $\alpha_{el}$,
  and the modulus of $Z$ vs $\Gamma$ calculated for the
  Hamiltonian in Eq.~(\ref{h0}) with $\delta=0$ for stacks
of 14 and 16 sites (circles and crosses, respectively).
  The arrows mark the NIT of the infinite strand.}
\label{flnalpha}
\end{figure}

In the present work, we take $\Psi(0)$ in Eq.~(\ref{psioff})
  to be the exact GS of $H_0(\delta, \Gamma)$. $P(0)$ in Eq.~(\ref{p})
  is the relevant GS dipole per site in units of $ea$.
Then, using $\Delta M(0)$
  in Eq.~(\ref{deltap}),
 we obtain $\Psi(F)$ exactly, compute $Z(F)$ and $P(F)$,
  and evaluate $\alpha_{el} = P'(0)$ numerically.
Figure \ref{flnalpha}a and b show, respectively, $\alpha_{el}$ and $Z(F)$
for $\delta = 0$ stacks of $N$ = 14 (circles) and 16 (crosses).
  To understand these results, we note that regular stacks have real $Z(0)$
  by symmetry and hence $P(0) = 0$ for any $N$.
We have $Z_x(0) \sim 1 $
at large positive $\Gamma$ where neutral diagrams with $q \sim 0$ dominate in
Eq.~(\ref{expect}) and $Z_x(0) \sim -1$ at large negative $\Gamma$ where
ionic diagrams with $q \sim \pm \pi$ dominate. $ Z_x(0)$
  vanishes at some $\Gamma \sim $ 0, as seen for both $N$ = 14 and 16.
The Jahn-Teller instability of $4n$ rings appears as a symmetry change of
$\Psi(0)$\cite{oldmixed,anu} where $Z(0)$ vanishes and changes
discontinuously in Fig. \ref{flnalpha}b.
The arrow at $\Gamma_c = -0.666$ marks the NIT of the extended regular stack as
found from extrapolations of symmetry crossovers.\cite{anu}
Finite-size effects in $Z(0)$ are reduced considerably in
$-N\ln(|Z(0)|^2)$, which is the proper size-independent
quantity.\cite{restaloc}
  The $\alpha_{el}$ results in Fig. \ref{flnalpha}a nearly coincide for $N$
  = 14 and 16 except around $Z \sim 0$.
In Section 5 we report almost identical $\alpha_{el}$ for $N$ = 14 and 16
  over the entire $\Gamma$
  range for stacks with equilibrium dimerization and $Z \ne 0$.
\begin{figure}
\begin{center}
\includegraphics* [scale=1] {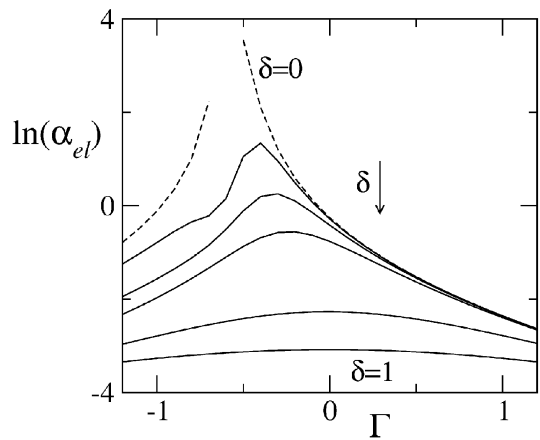}
\end{center}
\caption{Logarithm of the linear electronic  susceptibility, $\alpha_{el}$,
  vs $\Gamma$ for a 16 site stack with $\delta=$0.02, 0.05, 0.1, 0.4 and 0.99.
The arrows mark the direction of increasing $\delta$, and the dashed
line for $\delta=$ = 0 diverges at the NIT.}
\label{delta}
\end{figure}

         Figure \ref{delta} shows $\alpha_{el}(\delta,\Gamma)$ for N = 16.
Finite alternation, $\delta > 0$, strongly suppresses the divergence at the
NIT, as seen on comparing $\delta$ = 0.02 and 0 (dashed
line), and shifts the peak to $\Gamma = 0$ at $\delta =$ 1.
Regular stacks with a finite potential against dimerization have
  finite polarizability for $\Gamma \ge \Gamma_P > \Gamma_c$.
  The polarizability of dimerized stacks with $\Gamma  < \Gamma_P$
  remains finite because $\delta > 0$ opens a gap in the energy spectrum.
The $\delta$ = 0.99 curve follows closely the simple analytical result
for $\delta = 1$ and independent dimers,
\begin{equation}
\alpha_{el}=(\Gamma^2+8)^{-\frac{3}{2}}
   \label{dimers}
\end{equation}
when there is no electronic delocalization. The large variation of
$\alpha_{el}(\delta, \Gamma)$ in Fig. \ref{delta}
is understood in terms of localization around $\Gamma \sim 0$
  due to the dimerization gap and for $|\Gamma| >> 0$ due to
charge localization for any $\delta$. The pronounced asymmetry of
$\alpha_{el}(\delta,\Gamma)$ with respect to $\Gamma = 0$ is due to
strong correlations in $H_0(\delta, \Gamma)$ leading to quite
different GS on the neutral and ionic sides.

\begin{figure}
\begin{center}
\includegraphics* [scale=1] {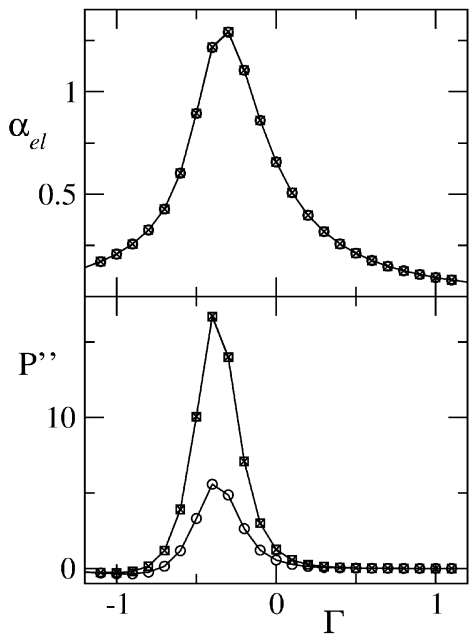}
\end{center}
\caption{The linear electronic susceptibility, $\alpha_{el}$,
and the second derivative of the
GS polarization on the electric field, $P''$, calculated
as a function of $\Gamma$ for a 16 site stack with $\delta= 0.1$. Circle,
squares and crosses refer to results obtained at the first second and
third iteration, respectively (see text).}
\label{itera}
\end{figure}
         We illustrate in Fig. \ref{itera}
  the iterative solution of Eq.~(\ref{hoff}).
  Circles refer to the first iteration with $\Delta M(0)$ for N = 16 and
$\delta = 0.1$. Crosses and squares refer to the second and third iteration,
respectively. The results for $\alpha_{el} = P'(0)$ do not change,
since the linear perturbation in $F$ appears from the outset and
iterations introduce higher-order corrections. Finite  $\delta$ is required
for finite $P''(0)$ and $F^2$ corrections must be included.
The second iteration more than doubles $P''(0)$, but the third iteration
  produces no additional change. Taylor expansion of $\Delta M(F)$
through the linear term suffices for $F^2$ contributions.
Nonlinear polarizabilities based on the PBC formulation in
Eq.~(\ref{energyfunctional}) or (\ref{hoff})
raise interesting issues for rigid lattices. Here we focus instead on
vibronic contributions in Eq.~(\ref{alphaelalphavib})
to the linear polarizability of models with Peierls transitions.

\section{ Polarizability near the Peierls transition }

  The Su-Schrieffer-Heeger (SSH) model \cite{ssh}
for polyacetylene, (CH)$_x$,
describes in the adiabatic approximation the Peierls instability of
H\"uckel chains with $U = \Delta = 0$ in Eq.~(\ref{hhu}) and linear
e-ph coupling $\alpha_{e-ph} = t\delta/u$. Similar approaches have been
applied to Peierls transitions in segregated ($\Delta = 0$) stacks
of $\pi$-electron donors or acceptors,\cite{ttftcnq}
  or to uncorrelated ($U = 0$) chains\cite{ricemele}
with site energies $\pm \Delta$.  The stability of mixed stacks described
by $H_0$ has been studied through the response of
a regular lattice to the SSH coupling.\cite{oldmixed}
In the adiabatic approximation, the GS potential energy surface is
defined by adding  the elastic energy for lattice motion
to the electronic energy. For harmonic potentials and linear
electron-phonon coupling, the elastic energy of the Peierls mode is
$\delta^2/2\epsilon_d$ per site, where $\epsilon_d=
\alpha_{e-ph}^2/k$ is the small polaron binding energy and $k$
is the lattice force constant. \cite{freo} We therefore define
the  Peierls-Hubbard
model by adding the lattice energy to $H_0(\delta, \Gamma)$ in
Eq.~(\ref{h0}). The electronic energy in an applied field is given by
Eq.~(\ref{energyfunctional}). The total GS energy
per site is
\begin{equation}
   {\cal E}_T= {\cal E}_0 (\delta, \Gamma, F) +\frac{\delta^2}{2\epsilon_d}
\label{etot}
\end{equation}
The same development holds for arbitrary $U, \Delta$ in Eq.~(\ref{hhu})
or for other cell models with PBC.

\begin{figure}
   \begin{center}
   \includegraphics* [scale=1] {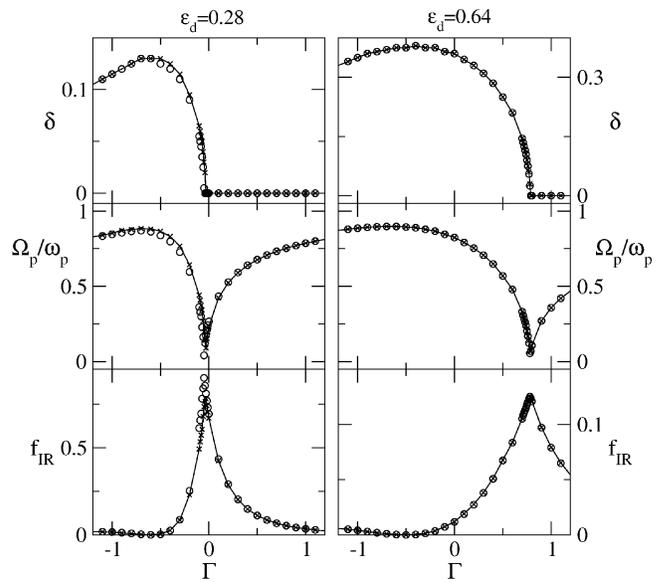}
   \end{center}
\caption{Equilibrium dimerization, $\delta$,
   softened frequency $\Omega_P$ in units of the reference
   frequency $\omega_P$, and infrared oscillator strength, calculated
   for Eq. (13) with $\epsilon_d=$ 0.28 and 0.64 (left and
   right panels, respectively.) Circles refer to $N$ = 14, crosses to
$N$ = 16.}
\label{v0}
\end{figure}

Eq. (\ref{etot}) fully defines the GS potential energy surface with
one additional parameter, the lattice stiffness $1/\epsilon_d$,
besides
those entering the electronic Hamiltonian. The equilibrium
dimerization is obtained by minimizing of the total energy,
\begin{equation}
   \delta_{eq}= -\epsilon_d \frac{\partial {\cal E}_0(\delta, \Gamma, F)}
   {\partial \delta}
\label{deltaeq}
\end{equation}
The limiting inverse stiffness is $\epsilon_d = \sqrt{2}$, which yields an
equilibrium dimerization in Eq.~(\ref{deltaeq}) of $\delta = 1$  at
$\Gamma = 0$ for a stack of decoupled dimers whose GS is easily found. The top panel of Fig. \ref{v0} shows the equilibrium dimerization of $H_0(\delta,\Gamma)$
  at $F = 0$ for $\epsilon_d =$ 0.28 and 0.64 in units of $t$. Since
results for N = 14 (crosses) and 16 (circles) now coincide almost
exactly, we conclude that finite-size effects have become negligible
in deformable lattices with $\delta > 0.1$ around $\Gamma \sim 0$ that precludes strong delocalization.

Previous modeling of TTF-CA used the estimate $\epsilon_d \sim 0.15-0.3$.
\cite{oldmixed,mixedparams}
The $\epsilon_d = 0.64$ stack is less stiff, with a three-fold increase in
the maximum dimerization to $\delta_{eq} \sim 0.3$ around $\Gamma \sim 0$.
The Peierls instability occurs at
$(\partial^2 {\cal E_T}/\partial \delta^2)_0=0$, where the curvature
$1/\epsilon_d$ of the lattice potential and the curvature of the electronic
energy cancel exactly. The stack is regular for
$\Gamma> \Gamma_P(\epsilon_d)$ and dimerized for
$\Gamma< \Gamma_P(\epsilon_d)$.
  As expected on general grounds, the neutral lattice with
$\Gamma >> 0$ is a band insulator that is conditionally stable.
The divergence of $(\partial^2 {\cal E}_0/\partial \delta^2)_0$
at $\Gamma_c$ ensures a Peierls instability at some
$\Gamma_P > \Gamma_c$ that is model dependent. Dimerization decreases but
persists as  $\Gamma$ becomes more negative.\cite{oldmixed}
  The $\Gamma << 0$ limit with
$\rho \sim 1$ maps into the spin-1/2 Heisenberg antiferromagnetic
chain\cite{mcconnell}
whose GS is unstable to a spin-Peierls transition.\cite{spin}

  The curvature of the GS potential energy surface is the frequency
of the Peierls mode
$\Omega_P^2=(\partial^2 {\cal E_T}/\partial \delta^2)_{eq}$.
At $F = 0$ the ratio $\Omega_P/\omega_P$ at the equilibrium $\delta$ is
a function of $\Gamma$:
\begin{equation}
   \left(\frac{\Omega_P}{\omega_P}\right)^2=1+\epsilon_d
\left(\frac{\partial^2{\cal E}_0(\delta, \Gamma)}{\partial \delta^2}
\right)_{eq}
\label{osuo}
\end{equation}
The middle panel of Fig. \ref{v0} shows the evolution of $\Omega_P/\omega_P$
  for the correlated model $H_0(\delta,\Gamma)$ in Eq.~(\ref{h0}).
  The softening of the Peierls mode on the neutral side and its subsequent
hardening are observed for TTF-QBrCl$_3$ whose continuous dimerization
transition has been investigated by far-IR spectroscopy.\cite{okimoto}
  The temperature evolution of a combination band in the mid-IR spectrum of
  TTF-CA for $T > T_c$, also supports the presence of a soft
lattice mode.\cite{combination}
The data are in both cases consistent with
  $\omega_P \sim$ 100 cm$^{-1}$, a typical frequency for lattice vibrations
in crystals with molecular masses of $\sim 100$ AMU.

  The Peierls mode is IR active by symmetry in mixed stacks
  and borrows huge IR intensity from electronic degrees of freedom.\cite{freo}
  In the present approximation, with $M$ in Eq.~(\ref{emme}) for fixed sites,
the IR intensity of the Peierls mode is entirely due to charge
fluctuations induced by $\delta$.
The corresponding oscillator strength is:
\begin{equation}
   f_{IR}=\frac{2m_e \Omega_P |\mu_{IR}|^2}{e^2 \hbar}=
\frac{m_e a^2\omega_P^2\epsilon_d}{t}
\left(\frac{\partial P}{\partial \delta}\right)_{eq}^2
\label{fir}
\end{equation}
Here $m_e$ and $e$ are the electronic mass and charge, respectively,
$|\mu_{IR}|^2$ is the squared transition dipole per site, and $\epsilon_d$
is in units of $t$.
Once again the IR intensity is governed by $P$-derivatives and $\epsilon_d$.
However   $\omega_P$, $a$ and $t$ enter the expression for the dimensionless
$f_{IR}$. The bottom panels of Fig. \ref{v0} show $f_{IR}$ for
typical parameters, \cite{mixedparams} $a$ = 3.7 \AA, $\omega_P$ = 100 cm$^{-1}$, $t$ = 0.2 eV.

The intensity of the Peierls mode has a pronounced peak at $\Gamma_P$
  where the electronic charges are maximally mobile.\cite{freo}
  Dimerization localizes charges and lowers $f_{IR}$ for $\Gamma < \Gamma_P$.
 We have $f_{IR}$ = 0 when the electronic flux induced by
dimerization reverses from D $\rightarrow$ A on the neutral side to
A $\rightarrow$ D on the ionic side. The neutral-ionic interface of
deformable lattices can be identified as $f_{IR} = 0$, which occurs near
$\Gamma = \Gamma_c$ for regular stacks or $Z_x(\delta, \Gamma) = 0$
  in dimerized stacks.\cite{freo}
  The actual IR intensity does not vanish due to
the motion of the molecular sites with charges $\pm e\rho$, but {\it frozen charge} contributions \cite{freo} are neglected in Eq.~(\ref{emme}).
In materials that dimerize in the neutral side,
the N-I interface could, in principle, appear experimentally
as a dip in $f_{IR}$. However, this dip does not mark a phase transition.
There is no change in symmetry and
all properties of the system vary continuously at the interface.
\cite{horovitz,oldmixed}

  The IR intensity of the stiff ($\epsilon_d$ = 0.28) stack in Fig. \ref{v0}
  is an order of magnitude larger than that of the $\epsilon_d$ = 0.64 stack.
While contrary to the $\epsilon_d$ factor in Eq.~(\ref{fir}),
this is readily understood in terms of greater delocalization in stiff
lattices whose $\Gamma_P$ approaches $\Gamma_c$ and whose
$\partial P/\partial \delta$ diverges in the limit $\epsilon_d
\rightarrow  0$. At the Peierls transition, the large oscillator strength
of the dimerization mode corresponds to effective masses
$m^* = m_e/ f_{IR} \sim$ 1000 and 6000 for $\epsilon_d$ = 0.28 and 0.64,
respectively. So $m^*$ at the transition is roughly a proton mass
and is fully two orders of magnitude smaller than
molecular masses.
This justifies {\it a posteriori} the neglect of frozen charge
  contributions to $f_{IR}$ near the Peierls transition.\cite{freo}

The small effective mass associated with dimerization at the Peierls
transition implies strong mixing of electronic and vibrational degrees
of freedom. Lattice contributions to the polarizability can be substantial,
as anticipated for $\alpha_{vib}$ in Eq.~(\ref{alphaelalphavib}).
To demonstrate this, we differentiate both sides of Eq.~(\ref{deltaeq})
  with respect to $F$ and use $P =-(\partial {\cal E}_0/\partial F)$ to obtain
  \begin{equation}
    \frac{\partial \delta_{eq}}{\partial F} =\frac{\epsilon_d}
{1+\epsilon_d(\partial^2 {\cal E}_0/\partial \delta^2)_{eq}}
\frac{\partial P}{\partial \delta}
\label{dddf}
  \end{equation}
We then use Eq.~(\ref{osuo}) to write the vibrational polarizability
of the Peierls mode at $\delta_{eq}$:
\begin{equation}
   \alpha_{vib}=\epsilon_d\left(\frac{\omega_P}{\Omega_P}\right)^2
\left(\frac{\partial P}{\partial \delta}\right)^2 \rightarrow
\frac{\epsilon_d \omega_P^2}{\Omega_P^2+\gamma^2}
\left(\frac{\partial P}{\partial \delta}\right)^2
\label{alphavib}
\end{equation}
This expression is exact when the vibrational kinetic energy is neglected,
which is a good approximation in view of the low frequencies
involved. \cite{ftp}
It is equivalent to the sum-over-states expression,\cite{ftp}
$\alpha_{vib}=2|\mu_{IR}|^2/\hbar \Omega_P$, based on IR transition moments
and energies.

\begin{figure}
\begin{center}
\includegraphics* [scale=1] {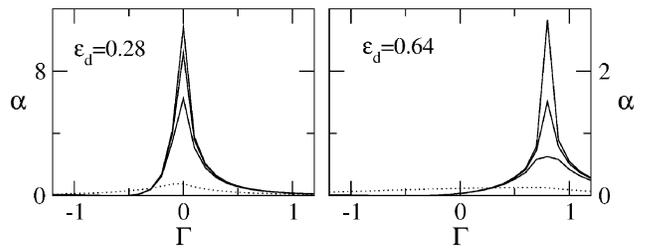}
\end{center}
\caption{Electronic and vibrational contributions to the static
  polarizability (dotted and continuous lines, respectively),
for $N = 16$ and the same parameters as in Fig. \ref{v0}. The three
lines for $\alpha_{vib}$ have, from top to bottom, for $\gamma/\omega_P=$ 0.01, 0.1, 0.2.}
\label{falphavib}
\end{figure}
  Large $\alpha_{vib}$ is expected for the dimerization mode due to
its large transition dipole and low frequency. Indeed,
$\Omega_P = 0$ at the Peierls transition gives a divergent polarizability.
We suppress the divergence by introducing damping $\gamma$
  that represents the lifetimes of lattice modes, anharmonic potentials, etc.
Damping is introduced empirically by changing
$(\Omega_P/\omega_P)^2$ as shown in the second equality in
Eq.~(\ref{alphavib}). Figure \ref{falphavib} reports $\alpha_{vib}$
for the $\epsilon_d$ = 0.28 and 0.64 systems shown in Fig. \ref{v0}
  for damping $\gamma/\omega_P$ = 0.01, 0.1 and 0.2.
  The dotted line is the electronic polarizability $P'(0)$ of the deformable
lattice. The indicated damping corresponds to $\gamma =$ 1, 10 and 20
cm$^{-1}$ for an estimated $\omega_P =$ 100 cm$^{-1}$.
  Quite predictably, damping affects $\alpha_{vib}$
  in a narrow region about $\Gamma_P$ and is more pronounced in the softer
  lattice. We note that typical bandwidths of lattice modes in molecular
  crystals are in the 1-10 cm$^{-1}$ range. We will use $\gamma = 0.1\omega_P$
to model the dielectric response of CT salts and cannot specify
$\alpha_{vib}$ peaks more accurately than shown in Fig. \ref{falphavib}.

\section{ Dielectric anomaly of Peierls-Hubbard models}

\begin{figure}
   \begin{center}
\includegraphics* [scale=1] {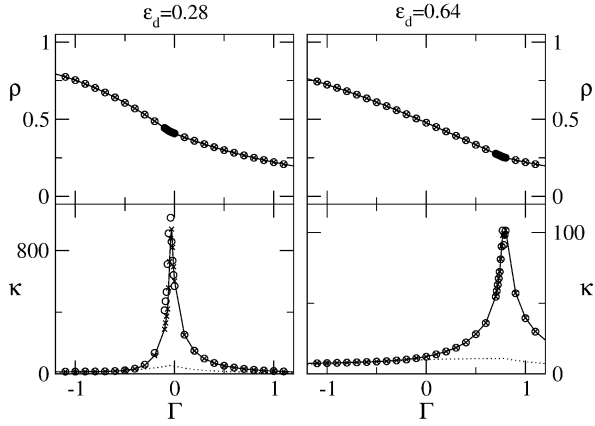}
\end{center}
\caption{The ionicity, $\rho$, and the dielectric response,
$\kappa$, calculated for the same parameters as in Fig. \ref{v0}.
Circles refer to N = 14, crosses to N = 16.
Dotted lines in the bottom panels show the dielectric response
obtained by neglecting the vibrational contribution to $\alpha$.}
\label{v0_1}
\end{figure}

         We now combine the electronic and vibrational
polarizabilities of the Peierls-Hubbard model in Eq.~(\ref{h0})
  to obtain the dielectric constant of the equilibrium lattice. In SI units,
  \begin{equation}
    \kappa=\kappa_\infty +\frac{\alpha_{el}+\alpha_{vib}}{\epsilon_0 v}
  \end{equation}
Here $\epsilon_0$ is the vacuum permittivity constant, $\kappa_\infty \sim 3$
  is the usual contribution from molecular excited states that are not
  being modeled, and $v$ is the volume per site.
We are interested in remarkably large  peaks with $\kappa > 100$.
  To get proper dimensions for $\kappa$ the polarizabilities ($\alpha_{el}$
and $\alpha_{vib}$) are multiplied by $e^2a^2/t$. We adopt
typical  TTF-CA values, $v$ = 206 \AA$^3$, $a$ = 3.7 \AA, and $t$ = 0.2 eV, and
in the lower panels of Fig. \ref{v0_1} we
show the resulting $\kappa = 3 + 60(\alpha_{el} +
\alpha_{vib})$ for $N$ = 14 and 16 stacks with equilibrium dimerization
and $\epsilon_d $= 0.28 and 0.64. The dotted lines are the electronic
contribution to $\kappa$.
The dielectric peak is clearly associated with the Peierls
transition and vibrations. The divergence of $\alpha_{el} $
in Fig. \ref{flnalpha}
  at the NIT ($\Gamma_c$ = -0.666) of the $\delta = 0$ stack is
strongly attenuated by dimerization and appears as a small bump around
$\Gamma_P$ in Fig. \ref{v0_1}.
  The electronic fluxes leading to the
  IR intensity of the Peierls mode are also
responsible for the large $\kappa$ peaks of $\sim $ 1000 and 100
estimated for $\epsilon_d = $
0.28 and 0.64, respectively. We note that neglecting the softening, i.e.
by imposing $\Omega_P = \omega_P$ in Eq.~(\ref{alphavib}), lowers
  the $\kappa$ peak by an order of magnitude, while neglecting
  damping ($\gamma = 0$) gives a divergence. We took $\gamma =
\omega_P/10 = 10$ cm$^{-1}$ in Fig. 6.

         Both the height and shape of the $\kappa$
  peaks as a function of temperature compare favorably to the dielectric data, discussed below, of Horiuchi et al. \cite{hprb,hprl,hjpsj,hjacs,okimoto}
That is not the case for the ionicity
$\rho$ shown in the upper panel of Fig. \ref{v0_1}. We properly have $\rho \sim $ 0.40 at the Peierls transition of the $\epsilon_d$ = 0.28 stack and
$\rho \sim $ 0.25 in the softer stack, but $\rho(\Gamma)$
  hardly shows any sign of a transition. By contrast,
the measured ionicity has a kink in systems with continuous transitions around $\rho \sim $ 0.25.\cite{hprb,hprl,hjpsj,hjacs,okimoto}
The basic model $H_0(\delta, \Gamma)$
  in Eq.~(\ref{h0}) is deficient in this respect, just as its
continuous NIT in rigid lattices cannot account for discontinuous
$\rho$ at $T_c \sim$ 81 K in TTF-CA. Several extensions of the model
have long been recognized. Coulomb interactions occur both within and between stacks.\cite{harding, strebel}
  Coupling to molecular modes is another general phenomenon.\cite{oldmixed}
  In practice, mean-field (mf) approximations
  are necessary for inter-stack interactions.
While observing a  $\rho$ anomaly at the Peierls transition
 requires going beyond $H_0(\delta,\Gamma)$,
the appropriate extensions and parameters remain open.

\begin{figure}
   \begin{center}
\includegraphics* [scale=1] {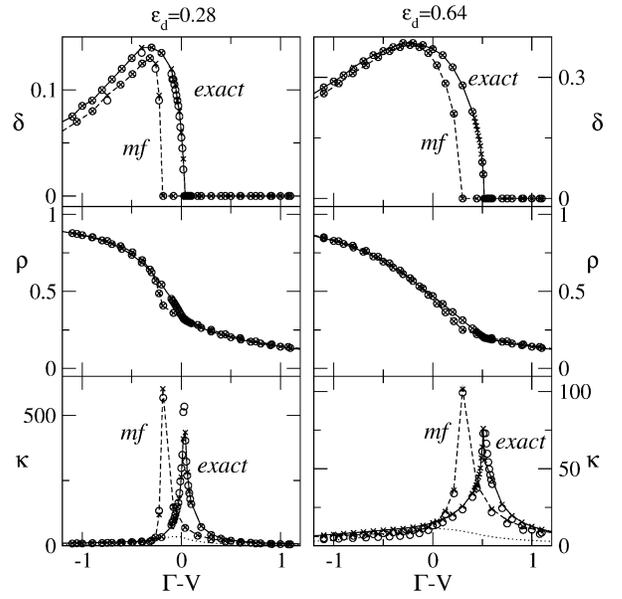}
\end{center}
\caption{The equilibrium dimerization amplitude, $\delta$,
the GS ionicity, $\rho$, and the dielectric response, $\kappa$,
for the same parameters as in Fig. \ref{v0} except for $V=2$. Circles and crosses refer to $N=$14 and 16, respectively, with
continuous and dashed lines joining exact and mf results.
Dotted lines in the bottom panels are the electronic mf
dielectric response for $N=16$.}
\label{v2}
\end{figure}
         To illustrate a simple extension that is suitable for either
exact or mf analysis, we introduce nearest-neighbor
Coulomb interactions $V$ along the stack,
\begin{equation}
   H_V=H_0(\delta,\Gamma)+V\sum_p q_p q_{p+1} +N\frac{\delta^2}{2\epsilon_d}
\label{hv}
\end{equation}
  Since $V$ is diagonal in the VB basis, all GS properties
are found as before. The solid lines in Fig. \ref{v2}
  are exact $N$ = 14 and 16 results for the equilibrium dimerization,
ionicity and dielectric anomaly of stacks with $V$ = 2 and otherwise
the same parameters as in Figs. \ref{v0} and \ref{v0_1}.
The dashed lines are the mf approximation to the $V$
  term in Eq.~(\ref{hv}). The mf results for $V$ = 2 are similar,
with somewhat sharper $\rho$ at $\Gamma_P$ as expected for an interaction
that stabilizes adjacent ion-pairs. Both $\epsilon_d$ = 0.28 and 0.64
now have a kink at $\rho(\Gamma_P)$ that qualitatively resembles
$\rho(T)$ data. The $\kappa$ peak  is lower by a factor of two,
while the maximum dimerization hardly changes. Charges are kept together for
$V > 0$, thereby reducing the polarizability.

\begin{figure}
   \begin{center}
\includegraphics* [scale=1] {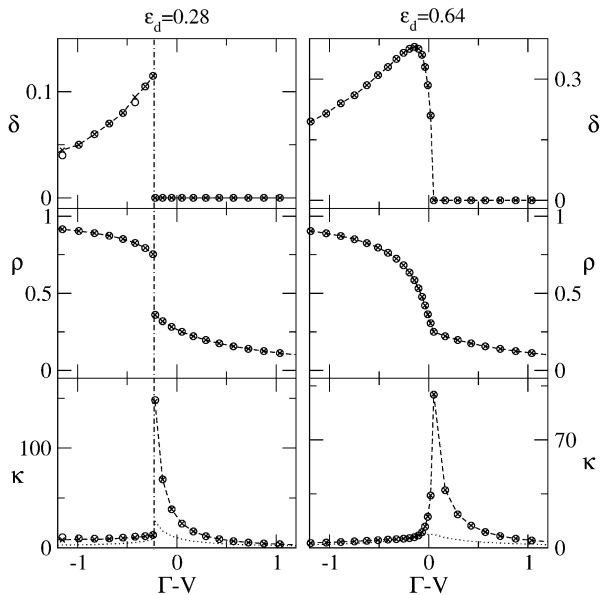}
\end{center}
\caption{The same as Fig. \ref{v2}, for $V=$ 3; for the sake of
clarity only mf results are reported, with circles and crosses referring
 to $N$ = 14, and 16, respectively. Dotted lines in the lowest panels show
the electronic contribution to $\kappa$.}
\label{v4}
\end{figure}
\begin{figure}
   \begin{center}
\includegraphics* [scale=1] {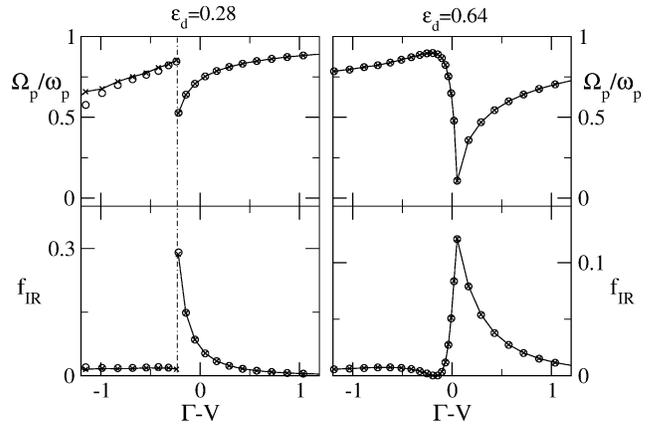}
\end{center}
\caption{Mean-field results for the softening of the Peierls mode and its
IR intensity, calculated for the same parameters as in Fig. \ref{v4}.
Circles refer to $N$ = 14, crosses to $N$ = 16.}
\label{vibv3}
\end{figure}

         The first-order transition of TTF-CA at 81 K has a
$\Delta \rho \sim 0.2$ and concomitant neutral-ionic and Peierls
transitions. \cite{ttfca,ttfcaj,ttfcar}
The ionicity is a suitable order parameter in this case. Discontinuous
$\rho$ has long been treated in mf theory,\cite{strebel} and mf results
for $V $= 3 are shown in Fig. \ref{v4} for the same quantities and parameters
as in Fig. \ref{v2}. Now
$\epsilon_d$ = 0.28 produces a $\Delta \rho$ jump in stacks of $N$
  = 14 and 16 that resemble previous results for smaller $N$ up to
12.\cite{oldmixed}
The kink in $\rho$ for $\epsilon_d$ = 0.64 is stronger and still larger $V$
  will make it discontinuous.
The maximum dimerization remains almost independent of $V$.
  The $\kappa$ peak now appears at the NIT and
is reduced for both $\epsilon_d$ = 0.28 and 0.64. The reason for
reduced vibrational contributions is that a discontinuous NIT
interrupts the softening of $\omega_P$. This is shown in Fig.
\ref{vibv3} for the same parameters, together with $f_{IR}$.

Figure \ref{v2} contrasts exact and mf results for small V = 2,
while Fig. \ref{v4} shows a first-order mf transition for V = 3
in the stiff stack. To summarize the various roles of $V$ in Eq.~(\ref{hv}),
we note that in mf it also accounts for the coupling of electrons with
molecular vibrations. Totally-symmetric
(ts) molecular vibrations modulate on-site energies,
and hence $\Gamma$, leading to Holstein coupling.\cite{rice}
The effects of this coupling on vibrational spectra of
CT salts are well known.\cite{rapid,bibbia,bozio}
In regular stacks, ts vibrations only appear in Raman spectra,
whereas in dimerized stacks they acquire large IR intensity,
proportional to $(\partial P/\partial \Gamma)^2$, through
their coupling to electronic degrees of freedom.\cite{iscom2002}
In the adiabatic approximation, Holstein coupling plays
exactly the same role as $V$ within mf.\cite{oldmixed,iscom2002}
Thus mf results in Figs. \ref{v2} and \ref{v4} represent an effective
parameter $V$ whose interpretation is model dependent.
Although the large IR intensity of Holstein modes in dimerized lattices
suggests that also they contribute to $\kappa$, their high frequency
($\sim $ 1000 cm$^{-1}$) makes this contribution negligible
in comparison to the Peierls modes.

\section{ Discussion}

         The dipole $P$ for insulators with PBC and its extension in Eq.~(\ref{energyfunctional})
to an applied field, make it possible to model
the electronic and vibrational polarizability of quantum cell models.
We found the induced-dipole operator $\Delta M$
in Section 2 for quantum cell models in general and applied it to a modified
  Hubbard model, $H_0(\delta, \Gamma)$ in Eq.~(\ref{h0}). We
report  model-exact electronic polarizabilities in Fig. \ref{delta}
  for the NIT of regular ($\delta$ = 0) and dimerized stacks.
We then include the Peierls mode with frequency $\omega_P$
  in the adiabatic approximation and its softening.
The IR intensity of the Peierls mode in Fig. \ref{v0} and the
dielectric peaks in Figs. \ref{v0_1} and \ref{v2} are related to
Peierls transitions in stacks with stiffness $1/\epsilon_d$.
We have chosen $\epsilon_d$ = 0.28 as a plausible estimate
  for TTF-CA and a softer
stack with $\epsilon_d $= 0.64 whose Peierls transition is around $\rho \sim$
  0.3 on the neutral side. The choice of $t$ = 0.2 eV comes from TTF-CA
optical data. The model parameters are not arbitrary, but typical of
values used for other properties. \cite{oldmixed,mixedparams} 
The same analysis holds for various
extensions of $H_0(\delta,\Gamma)$ for organic CT
salts. All  results are derived from the $\Gamma$-dependence of the
 GS properties of stacks described by  $H_0(\delta,\Gamma)$, and at the
 equilibrium dimerization for fixed
$\epsilon_d$ in Eq.~(\ref{etot}) and fixed nearest-neighbor interactions
$V$ in Eq.~(\ref{hv}).

         The dielectric data of Horiuchi et al. are taken as a function of
temperature at ambient pressure,\cite{hprb,hprl,hjpsj,hjacs,okimoto}
   or, more recently, as a function of pressure, up to  $\sim$ 10 kbar,
at constant temperature. \cite{science}
Decreasing volume on cooling or compression
stabilizes the ionic phase by increasing the crystal's Madelung energy.
Cooling corresponds to decreasing $\Gamma$  and increasing $\rho$
in models, but the relation between temperature  or pressure and
$\Gamma$ is not known at present.
Cooling also increases DA overlaps that enter in $t$
  and $\epsilon_d$ for one-dimensional models. Overlap between
stacks also increases, but although the importance of inter-stack
interactions has been noted,\cite{ttfcar} their modeling
  is still rather qualitative.

         The crystals studied by Horiuchi et al.
in refs. \onlinecite{hjacs}, \onlinecite{science} have D = DMTTF
(dimethyl-TTF) and the quinones A = QBr$_n$Cl$_{4 - n}$ with $n$ = 0-4;
$n$ = 0 and 4 are CA and bromanil (BA), respectively.
The acceptors have two-fold disorder in Br/Cl in the crystal lattice for $n$
  = 1, 2 and 3. There are two QBr$_2$Cl$_2$ isomers, a centrosymmetric
one called 2,5 and a polar one called 2,6. The $n$ = 1, 3 and 2,6
complexes are formally polar, but only weakly so: Their dielectric
peaks are similar but are shifted to lower temperature.
As there is no disorder in the prototypical system, TTF-CA, the best
candidates with continuous transitions are DMTTF-CA and DMTTF-2,5QBr$_2$Cl$_2$.
  The dielectric peak of DMTTF-CA at 65 K is $\kappa \sim$ 220 at 30 kHz;
$\rho$ increases quickly from $\sim $ 0.32 to $\sim $ 0.42 between 65 and 60 K
and then slowly to $\sim $ 0.48 at 10 K. The DMTTF-2,5QBr$_2$Cl$_2$
  peak at 50 K is $\kappa \sim$ 170 at 30 kHz, with $\rho$
  increasing from 0.29 to 0.36 on cooling from 50 to 10 K. The $\kappa$
peak of DMTTF-QBrCl$_3$ at $\sim$ 55 K closely resembles the 2,5 system,
but $\rho$ has not been reported. Pressure-induced $\kappa$
  peaks are reported \cite {science} at 5 K in DMTTF-BA ($n$ = 4) and in DMTTF-2,6QBr$_2$Cl$_2$, both with $\rho <$ 0.3 at 5 K and 1 atm.

         The observed peak heights of $\kappa$
  are consistent with dielectric peaks of the model
at the Peierls transition. We conclude that a $\kappa$ peak marks the
Peierls transition in systems with continuous $\rho$. The narrower $\kappa$
  peak in Fig. \ref{v4} for model parameters leading to a first-order
transition is also seen in the narrow $\kappa = 600 $ peak of TTF-CA.
\cite{hjpsj} The peak shapes and heights follow qualitatively the calculated
pattern of increased polarizability in stiff lattices
with Peierls transitions (continuous $\rho$) occurring at $\Gamma_P$
  close to $\Gamma_c$. More quantitative analysis requires a relation between
temperature and $\Gamma$. As illustrated in Section 5, accurate modeling of
$\rho(\Gamma)$  near $\Gamma_P$ depends on such specifics as damping and
intra and inter-stack interactions.
  By contrast, the Peierls transition of deformable stacks and NIT
  of rigid stacks are general features of Hubbard and related models.

The dielectric anomaly observed in
mixed stack CT salts is clearly associated with lattice degrees of freedom
and the Peierls transition in systems with either continuous or discontinuous
NIT. The huge IR intensity of the Peierls mode around the Peierls transition\cite{freo} and its softening
are responsible for $\alpha_{vib}>>\alpha_{el}$ in
Eq.~(\ref{alphaelalphavib}). Available spectroscopic data\cite{okimoto}
 in the far-IR
region for TTF-QBrCl$_3$ support this picture.\cite{freo}
It is interesting to compare the Peierls transition occurring in this system at
$\rho \sim 0.3$ with the discontinuous NIT occurring in TTF-CA at similar
$\rho$:  in TTF-CA the NIT occurs before the complete softening of the
Peierls mode and a reduced $\kappa$-peak is observed.
Whereas far-IR data are not available for TTF-CA, the incipient softening of
the dimerization mode has been recently extracted from a detailed study of
combination bands in the mid-IR region.\cite{combination}
A more systematic analysis of far-IR spectra and dielectric properties
of mixed stack materials is certainly desirable as it will confirm the
important connection between lattice degrees of freedom and materials
properties.

Organic CT crystals have DA repeat units along the stack and,
in contrast to most inorganic salts, are quasi-one-dimensional systems.
Rapid convergence, often as $1/N^2$, is typical for GS properties of
systems with PBC and small repeat units.
It is very convenient for modeling that $P(F)$ is a GS property.
 As found throughout, all $N = 14$ and $N = 16$ results coincide
in deformable lattices with $\delta > 0.1$ at $\Gamma \sim 0$.
 Larger N is not required for CT salts, but is needed for the
NIT of $\delta = 0$ stacks.

Discontinuous $\rho$ at the NIT is fundamentally different
from continuous $\rho$ at the Peierls transition of softer lattices.
We have focused on the Peierls-Hubbard model, $H_0(\delta,\Gamma)$ in
Eq.~(\ref{h0}), which has continuous $\rho$
  and a Peierls instability on the neutral side
for large $\epsilon_d$. The order parameter
$\delta$ for the second-order transition breaks the reflection symmetry
of the regular chain. Once the stack has dimerized, there is no
further transition and, indeed, hardly any remnants of the prominent
  neutral-ionic transition of the $\delta = 0$ stack.
Models with continuous $\rho$
  have Peierls rather than neutral-ionic transitions.
Similar results are obtained for stacks with not too large $V$-like
interactions: The Peierls or structural transition occurs at small $\rho$, and $V$ promotes a more prominent kink in $\rho$. Increasing further
$V$ (or Holstein) interactions leads to a discontinuous NIT, where the on-site
charge reorganization is the driving event, $\rho$ is the proper
order parameter, and dimerization just follows from the unconditional
instability of the ionic phase.

The increasing number of CT salts undergoing phase
 transitions demonstrates\cite{ricememorial}
 far more diversity than the neutral-regular and ionic-dimerized GS of TTF-CA. The lattice stiffness $1/\epsilon_d$ is an important parameter for distinguishing between Peierls and neutral-ionic transitions. Disordered D or A in single crystals raises different issues. The microscopic modeling of CT salts has primarily been in terms of Hubbard-type models with many open questions about parameters. We anticipate that accurate modeling of $P(F)$ and its applications to dielectric, vibrational and other data will lead to more quantitative descriptions of quantum phase transitions in these materials.

\begin{acknowledgments}
Z.S.  thanks R.Resta for fruitful discussions, and A.P.
 likewise thanks S. Ramasesha and F. Terenziani. Work in Parma was
partly supported by the Italian Ministry of Education (MIUR) through
COFIN-2001, and by INSTM through PRISMA-2002.
\end{acknowledgments}

\end{document}